\documentclass[aip,jcp,graphicx]{revtex4-1}
\usepackage[version=3]{mhchem} 
\usepackage{graphicx}

\draft 

\begin{document}


\title{Bias induced spin state transition mediated by electron excitations} 



\author{Hua Hao}
\email[]{huahao@pitt.edu}
\affiliation{Department of Chemistry, University of Pittsburgh, 219 Parkman Avenue, Pittsburgh, Pennsylvania 15260, United States.}
\affiliation{Key Laboratory of Materials Physics, Institute of Solid State Physics, Chinese Academy of Sciences, Hefei 230031, China.}

\author{Ting Jia}
\affiliation{Key Laboratory of Materials Physics, Institute of Solid State Physics, Chinese Academy of Sciences, Hefei 230031, China.}

\author{Xiaohong Zheng}
\affiliation{Key Laboratory of Materials Physics, Institute of Solid State Physics, Chinese Academy of Sciences, Hefei 230031, China.}

\author{Peng Liu}
\affiliation{Department of Chemistry, University of Pittsburgh, 219 Parkman Avenue, Pittsburgh, Pennsylvania 15260, United States.}
\email[]{pengliu@pitt.edu}

\author{Zhi Zeng}
\email[]{zzeng@theory.issp.ac.cn}
\affiliation{Key Laboratory of Materials Physics, Institute of Solid State Physics, Chinese Academy of Sciences, Hefei 230031, China.}




\begin{abstract}
Recent experiments reported that spin-state transitions were realized by applying bias voltages. But these bias-induced spin state transitions (BISSTs) are not fully understood, especially the mechanism. It is well known that the metal-to-ligand charge transfer excitation (MLCT) and the metal-centered excitation (MC) activated by light radiation can induce the transition from low spin (LS) to high spin (HS) and that from HS to LS. Moreover, electronic excitations are accessible by inelastic cotunneling in molecular junctions with bias voltages applied. Based on these two experimental facts, we propose the MLCT basically leads to the BISST from LS to HS, and the MC results in the BISST from HS to LS. The rationality of the mechanism is demonstrated by comparing first-principles results and experimental observations. The calculated voltage threshold for activating the MLCT (MC) is close to the experimental voltage for observing the BISST from LS to HS (from HS to LS). The activation of MLCT (MC) depends on the bias polarity, which can explain the bias-polarity dependence of BISST in the experiment. Our study is important for further design of molecular spintronic devices working on spin state transition.
\end{abstract}


\maketitle 

\section{Introduction}

On the molecular scale manipulation of spin by electrical means is attractive and important, especially for molecular spintronic devices\cite{Sanvito2011,Rocha2005,Bogani2008}. Such a manipulation is simpler and of higher spatial resolution than that by other means (e.g. temperature, pressure, light and magnetic field), which helps to simplify the architecture of devices and increase the density of functional units. Recent experiments reported that spin crossover molecules can be used to achieve the spin state transition (SST) by applying bias voltages in molecular junctions\cite{Prins2011,Miyamachi2012,Gopakumar2012}. The involved molecules have a nearly octahedral structure in which one ferrous ion of $3d^{6}$ is surrounded by several organic ligands. The ground (metastable) state is related to a low-spin (high-spin) metal ion\cite{Gutlich1994,Bousseksou2011}. The switch between the low-spin ground state and the high-spin metastable state is accompanied by a great change in electron transport property\cite{Huang2016,Aravena2012,Baadji2012} and optics (a pronounced, visible change of colour)\cite{Kahn1998}. Consequently, spin crossover molecules and the related bias-induced SST (or BISST) have greatly potential applications in molecular switches, molecular transistors, molecular memory devices, display devices, etc\cite{Letard2004}.

In spite of the above achievements, the BISST is not fully understood, especially its mechanism. Several possibilities were suggested to explain the BISST from low spin (LS) to high spin (HS). For example, local resistive heating or the Joule effect (due to inelastic scattering of tunneling electrons) may trigger the BISST from LS to HS\cite{Miyamachi2012,Prins2011}. This BISST may be caused by the Stark effect\cite{Diefenbach2007,Baadji2009,Hao2012}. However, it is observed that the BISST from LS to HS depends on the bias polarity in experiments, that is, it can be achieved only by applying negative bias voltages to the molecular junction but not by positive bias voltages\cite{Miyamachi2012}, or vice versa\cite{Gopakumar2012}. In view of this feature, the Joule-heat mechanism is hardly responsible for the BISST from LS to HS because the inelastic scattering of tunneling electrons is present at both bias polarities. For the similar reason, the Stark effect is not the cause for the BISST as well, which has been confirmed by the experiment\cite{Gopakumar2012}. In addition, our previous work\cite{Hao2017} suggested the BISST from LS to HS stimulated by one electron added to the molecule. Based on this mechanism, one can comprehend the BISST dependent on the bias polarity, but the voltage threshold is much larger than the experimental voltage for observing the BISST\cite{Miyamachi2012}. As for the BISST from HS to LS, \citet{Miyamachi2012} guessed that it is closely related to the $d$-$d$ transition, but this mechanism has not been confirmed until now, and its dependence on the bias polarity has been also unclear.

In the present work, we propose that the metal-to-ligand charge transfer excitation (MLCT) basically causes the BISST from LS to HS, and the metal-centered excitation (MC) or $d$-$d$ transition triggers the BISST from HS to LS. Our proposed mechanism is based on two well-established facts in experiments. One is that the light-induced MLCT and MC can generate the SST from LS to HS\cite{Chergui2013} and that from HS to LS\cite{Hauser1995,Gutlich2000}, which is commonly called the light-induced excited spin state trapping (LIESST). The other is that electronic excitations including MLCT and MC can be stimulated by the inelastic electron cotunneling through molecules, which is a two-electron tunneling process in the Coulomb blockade region\cite{DeFranceschi2001,Osorio2008}. To show the rationality of our proposed mechanism, time-dependent density-functional theory (TD-DFT) calculations are performed and the molecule \ce{Fe(phen)2(NCS)2} (\ce{phen}=1,10-phenanthroline) is taken as example. The calculated voltage threshold for activating MLCT (MC) is close to the bias voltage used to observe the BISST from LS to HS (from HS to LS) in the experiment\cite{Miyamachi2012}. Moreover, it is revealed that the activation of MLCT (MC) is determined by the bias polarity, which can explain the bias-polarity dependence of BISST from LS to HS (from HS to LS).

\section{Computational details}

The spin crossover molecule \ce{Fe(phen)2(NCS)2} is taken as example because it is involved in a scanning tunneling microscopy (STM) experiment\cite{Miyamachi2012}, where the bias-polarity dependence of BISST is observed and bias voltages used to observe the BISST between LS and HS are well recorded. The other optical experiment based on the same molecule shows the MLCT is responsible for the transition from LS to HS, and the MLCT (MC) energy is recognized\cite{Bertoni2015}. These help to verify our calculation method and the rationality of our proposed mechanism. In \ce{Fe(phen)2(NCS)2}, the ferrous ion with $3d^6$ is in the center of a nearly octahedral structure and surrounded by two different kinds of ligands (\ce{NCS} and \ce{phen}). $3d$ orbitals of the ferrous ion are split into two higher-level \ce{e_g} and three lower-level \ce{t_{2g}} orbitals. The electron configuration in the high-spin state is \ce{t^4_{2g}}\ce{e^2_g}, and that in the low-spin state is \ce{t^6_{2g}}\ce{e^0_g}. Our study is based on the free or gas-phase molecule since the couplings of the molecule with the tip and substrate are generally quite weak in experiments\cite{Miyamachi2012,Prins2011,Gopakumar2012}, and electronic structures are little affected by metallic electrodes\cite{Gueddida2013}. To optimize the structure, the B3LYP functional along with the 6-31G(d) basis set is employed in unrestricted open-shell calculations. This combination usually yields reliable ground-state geometries. We use the PBE0 functional\cite{Adamo1999} and the 6-311G(d,p) basis set to calculate excitation energies of the considered molecule, and results agree well with optical observations\cite{Bertoni2015,De2018,Glijer2008,Buron2012,Hauser1986,Hauser1991,Romstedt1998}. The spin multiplicity of the excited state is the same as that of the ground state for both high-spin and low-spin molecules. All calculations are performed using Gaussian09\cite{G09}.

\begin{figure}
  \centering
  \includegraphics[scale=0.8]{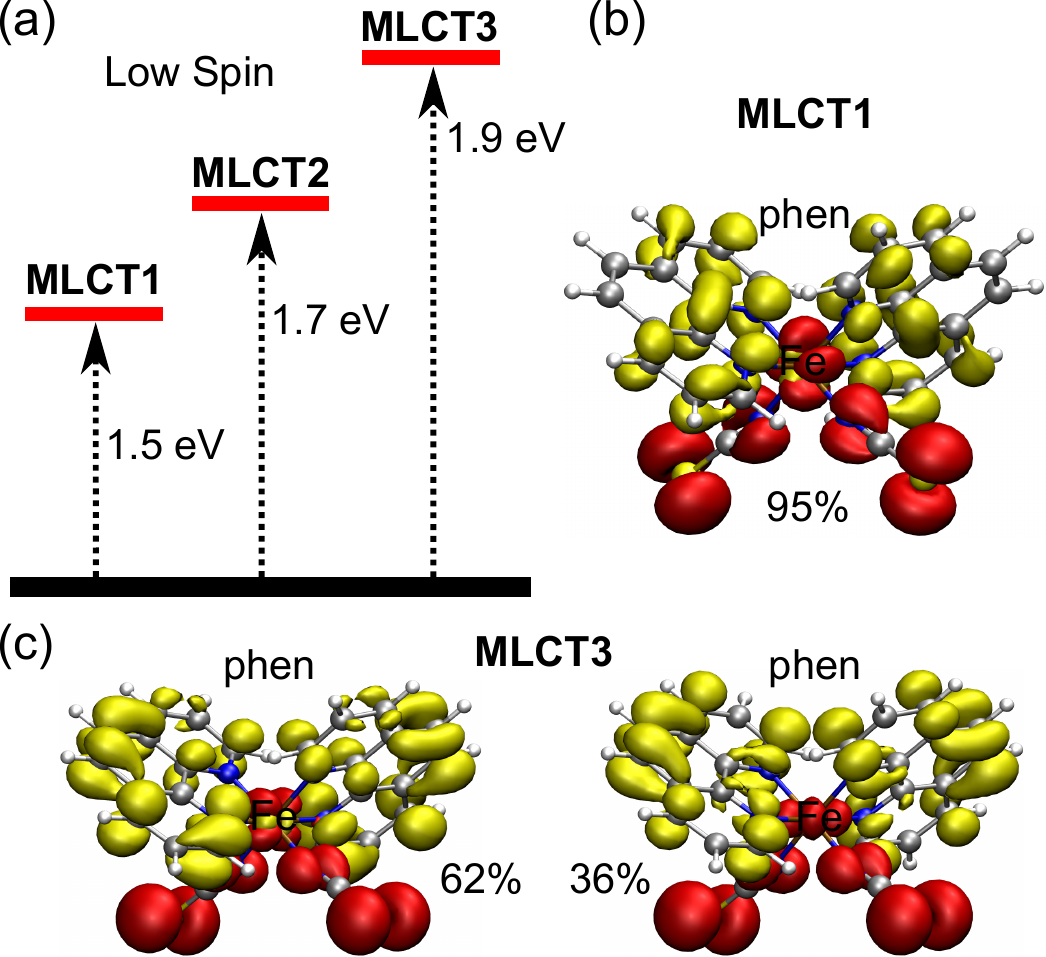}
  \caption{\label{fig1} (color online). (a) Energies of three lowest metal-to-ligand charge transfer excitations (\textbf{MLCT1-3}) in the low-spin state. (b) and (c) Natural transition orbital (NTO) analysis of \textbf{MLCT1} and \textbf{MLCT3}. The hole NTO is marked by red isosurface and yellow isosurface denotes the particle NTO. For \textbf{MLCT1-3}, the electron is generally excited from a \ce{t_{2g}}-like orbital to a p-character orbital of the (\ce{phen}) ligand. The percentage shows the proportion of the related hole-particle transition in the overall excitation.}
\end{figure}

\section{Results and discussion}

Firstly, we focus on the MLCT in the low-spin state. The calculated voltage threshold for stimulating the lowest MLCT is found to be consistent with bias voltages inducing the BISST from LS to HS in the experiment\cite{Miyamachi2012}. Besides, the realization of MLCT is affected by the bias polarity, which can explain why the BISST from LS to HS is observed only at negative bias in the experiment. These two results support the lowest MLCT leading to the BISST from LS to HS. Details are in the following.

Fig. \ref{fig1} (a) shows the lowest MLCT (\textbf{MLCT1}) is $1.5$ eV higher than the ground state, and the other two MLCTs are $1.7$ eV (\textbf{MLCT2}) and $1.9$ eV (\textbf{MLCT3}) higher. Features of these three MLCTs are analyzed using the natural transition orbital (NTO) method\cite{Martin2003}. Taking \textbf{MLCT1} as an example, the hole NTO has an obvious \ce{t_{2g}} character of the metal ion and the particle NTO is characterized by $p_z$ orbitals of two (\ce{phen}) ligands, as displayed in Fig. \ref{fig1} (b). According to the NTO theory, an electronic excitation is depicted by transitions from hole NTOs to particle NTOs, thus this excitation is regarded as a MLCT. The percentage ($95\%$) indicates that \textbf{MLCT1} is determined by one pair of hole and particle NTOs. The MLCT feature is similarly present at \textbf{MLCT2} and \textbf{MLCT3}, but there are two pairs of hole and particle NTOs contributing to them [see Fig. \ref{fig1} (c)]. The NTOs of \textbf{MLCT2} are not shown here. Our calculated energy of \textbf{MLCT3} is consistent with the optical observation\cite{Bertoni2015}, where a MLCT of \ce{Fe(phen)2(NCS)2} is detected by the pump light at $650$ nm ($1.9$ eV). Other optical experiments\cite{De2018,Glijer2008,Buron2012} demonstrate that a MLCT can be attained by an infrared radiation (around $800$ nm or $1.5$ eV), close to the calculated energy of \textbf{MLCT1}. Hence, our calculation based on PBE0/6-311G(d,p) is reliable to study the MLCT of the considered molecule.

\begin{figure*}
  \includegraphics[scale=0.8]{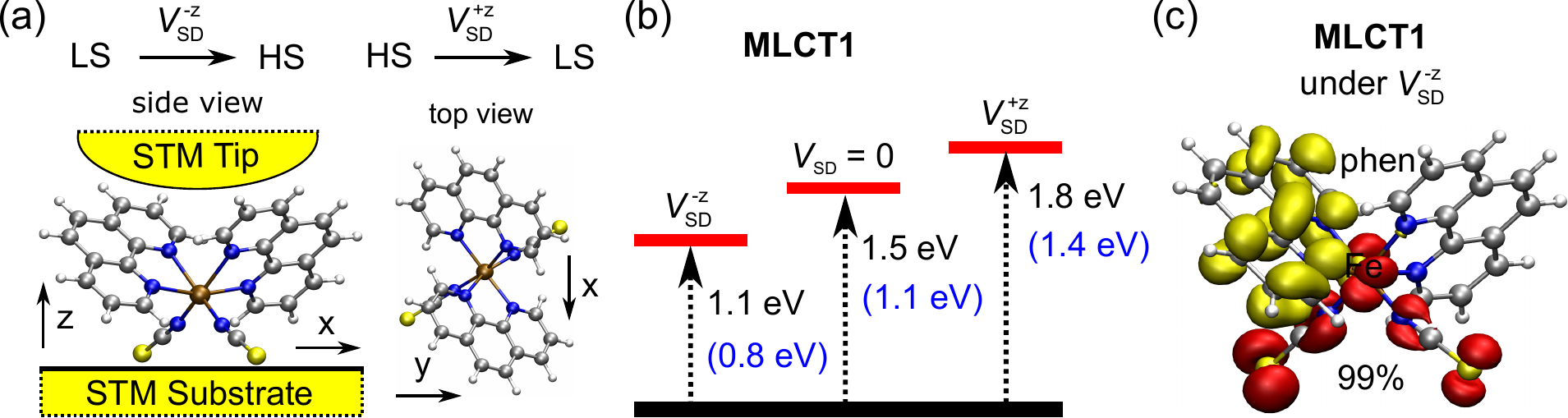}
  \caption{\label{fig2} (color online). (a) Schematic drawing of the molecular junction in the STM experiment\cite{Miyamachi2012} and the spin-state transition dependent on the bias polarity. $V^{\texttt{+z}}_{\texttt{SD}}$ and $V^{\texttt{-z}}_{\texttt{SD}}$ denote positive and negative bias voltages applied in this experiment. The electrostatic field induced by $V^{\texttt{+z}}_{\texttt{SD}}$ is along the z direction and opposite to the $V^{\texttt{-z}}_{\texttt{SD}}$ induced field. (b) Energies of \textbf{MLCT1} after considering different tunneling conditions. Values without parenthesis are based on the ground-state structure, and those in parenthesis based on the relaxed \textbf{MLCT1} structure. (c) NTO analysis of the relaxed \textbf{MLCT1} under $V^{\texttt{-z}}_{\texttt{SD}}$.}
\end{figure*}

To accurately depict the MLCT in the BISST experiment\cite{Miyamachi2012}, it is necessary to consider the effect of couplings with the STM substrate and that of bias voltages on excitations. As for couplings with the STM substrate, Gueddida and Alouani has clearly demonstrated when the low-spin molecule is absorbed on the STM substrate [CuN/Cu(001)] in the experiment, its electronic structure is almost the same as that in the free case\cite{Gueddida2013}. Thus for simplicity, the free molecule is used to simulate excitations in the molecular junction. Additionally, the BISST from LS to HS is observed by applying negative bias voltages to the substrate but not by positive bias voltages in the experiment, schematically displayed in Fig. \ref{fig2} (a). To consider the effect of the negative bias voltage, we apply an uniform electrostatic field with a moderate magnitude of $0.2$ V/\AA\ (marked by $\vec{E}^{\texttt{-z}}_{\texttt{UF}}$) to the molecule, and $\vec{E}^{\texttt{-z}}_{\texttt{UF}}$ is along the direction opposite to \texttt{z} in Fig. \ref{fig2} (a). Then Fig. \ref{fig2} (b) shows the energy of \textbf{MLCT1} is decreased from $1.5$ to $1.1$ eV with $\vec{E}^{\texttt{-z}}_{\texttt{UF}}$ applied. The lowered energy of \textbf{MLCT1} is because the electric dipole moment induced by the hole-particle transition in Fig. \ref{fig1} (b) (or transition dipole moment) has the same direction with the electrostatic field and \textbf{MLCT1} is favored by the negative bias. It has been known that an excitation can be achieved by inelastic cotunneling only if $eV_{\texttt{SD}} \geq \Delta E_{\texttt{e}}$ is satisfied\cite{DeFranceschi2001,Osorio2008}, where $V_{\texttt{SD}}$ is the applied bias voltage and $\Delta E_{\texttt{e}}$ is the energy of this excitation relative to the ground state. Accordingly, the voltage threshold for activating \textbf{MLCT1} is $1.1$ V in this case, which coincides with the experiment that the BISST from LS to HS can be observed by applying a negative bias voltage of $1.2$ V. Moreover, the negative bias voltage inducing the BISST from LS to HS can further decrease to $0.8$ V in the experiment. This bias voltage is probably because of the structural relaxation of \textbf{MLCT1}. After considering both the structural relaxation and $\vec{E}^{\texttt{-z}}_{\texttt{UF}}$, our calculated energy of \textbf{MLCT1} is also reduced to $0.8$ eV [marked by parenthesis in Fig. \ref{fig2} (b)], which is still consistent with the experiment. The NTO analysis in Fig. \ref{fig2} (c) exhibits the MLCT feature of \textbf{MLCT1} when these two factors are taken into account. In summary, the calculated voltage threshold for activating \textbf{MLCT1} agrees well with bias voltages used to observe the BISST from LS to HS in the experiment.

\begin{figure*}
  \includegraphics[scale=0.8]{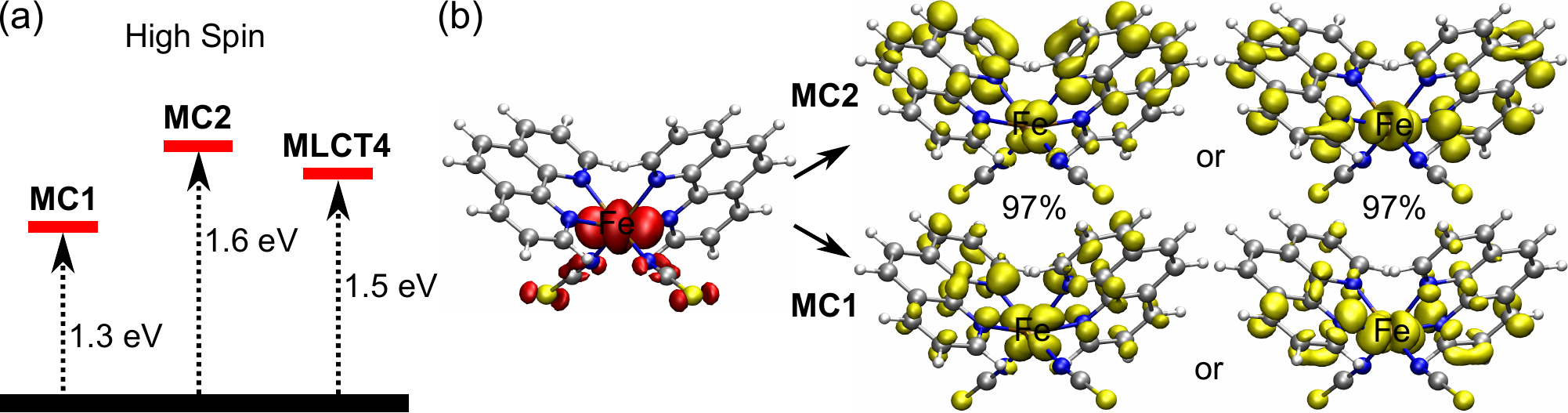}
  \caption{(color online). (a) Energies of two lowest metal-centered excitations (\textbf{MC1-2}) and one lowest MLCT (\textbf{MLCT4}) in the high-spin state, calculated based on the isolated molecule. (b) NTO analysis of \textbf{MC1} and \textbf{MC2}, where the electron is generally excited from a \ce{t_{2g}}-like orbital to an \ce{e_{g}}-like orbital.}
  \label{fig3}
\end{figure*}

\begin{figure}
  \includegraphics[scale=0.8]{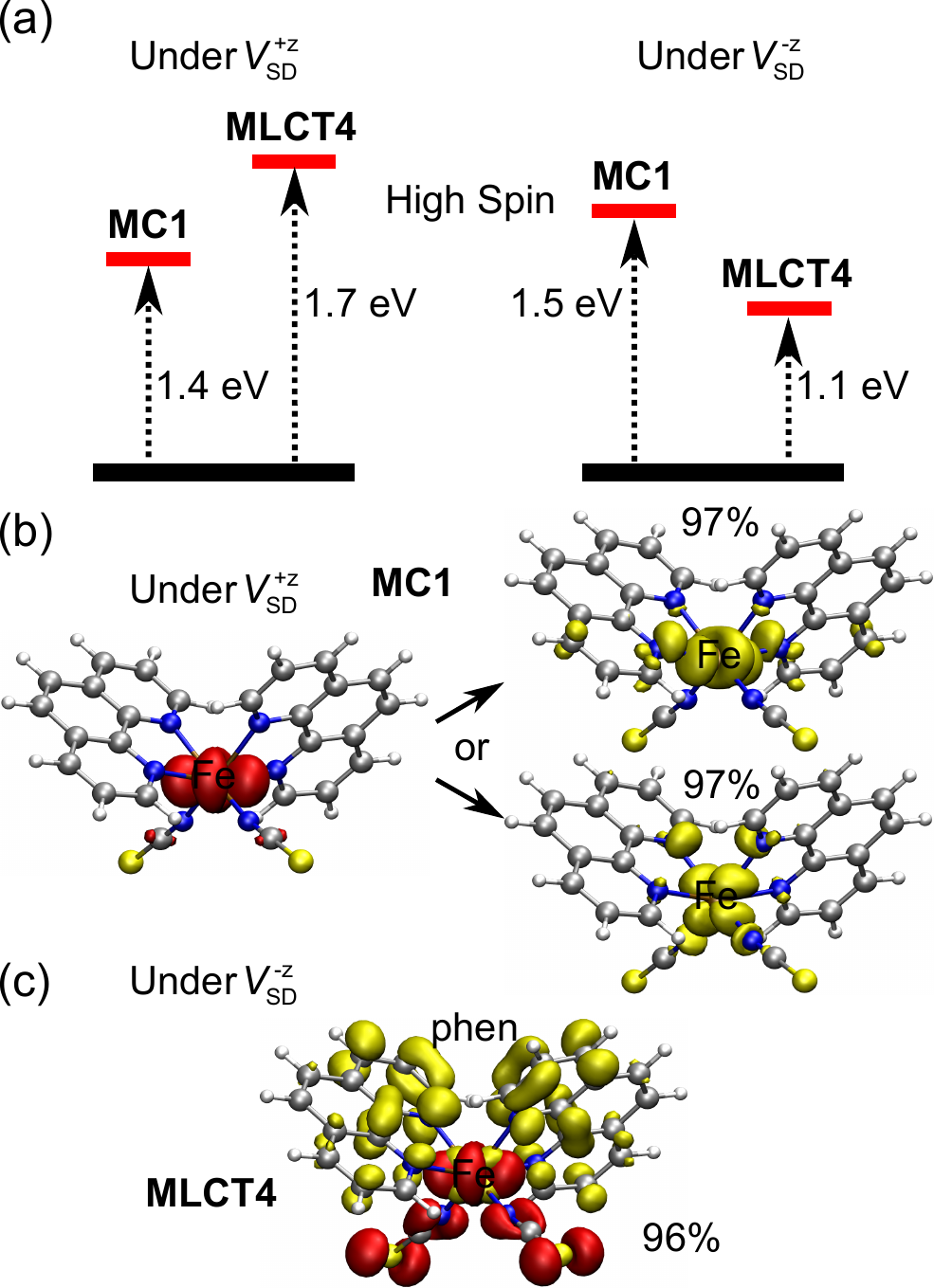}
  \caption{(color online). (a) Energies of \textbf{MC1} and \textbf{MLCT4} after considering different tunneling conditions. (b) NTO analysis of \textbf{MC1} under positive bias ($V^{\texttt{+z}}_{\texttt{SD}}$) and that of \textbf{MLCT4} under negative bias ($V^{\texttt{-z}}_{\texttt{SD}}$).}
  \label{fig4}
\end{figure}

Here, we will illustrate \textbf{MLCT1} becomes unaccessible at positive bias. For this purpose, we calculate the energy of \textbf{MLCT1} under $\vec{E}^{\texttt{+z}}_{\texttt{UF}}$ with the magnitude of $0.2$ V/\AA. $\vec{E}^{\texttt{+z}}_{\texttt{UF}}$ is used to simulate the effect of the positive bias voltage, and is along the \texttt{z} direction [see Fig. \ref{fig2} (a)]. Based on this model, the energy of \textbf{MLCT1} is increased from $1.5$ to $1.8$ eV [Fig. \ref{fig2} (b)], and the energy is still as high as $1.4$ eV even further including the structural relaxation. Owing to the increased energy, \textbf{MLCT1} becomes hardly accessible by the positive bias voltage ($1.2$ V) in the experiment, and the transition from LS to HS mediated by MLCT is not observable. This result is consistent with the absent BISST from LS to HS at positive bias in the experiment.

Secondly, we focus on the MC or $d-d$ transition in the high-spin state. The calculation illustrates the voltage threshold for activating the MC is consistent with bias voltages inducing the BISST from HS to LS in the experiment\cite{Miyamachi2012}. Moreover, the MC is prominently observed only at positive bias, which can explain why the BISST from HS to LS is achieved only at positive bias in the experiment. These results support the MC or $d-d$ transition responsible for the BISST from HS to LS.

Fig. \ref{fig3} (a) shows the two lowest MC levels (\textbf{MC1} and \textbf{MC2}) are respectively $1.3$ and $1.6$ eV higher than the ground state. For each MC level, there are two degenerate excitations, which are from the same \ce{t_{2g}}-like orbital to two different \ce{e_g}-like orbitals [see Fig. \ref{fig3} (b)]. The energy difference between \textbf{MC1} and \textbf{MC2} is due to different orbitals of the \ce{phen} ligands mixed into $d-d$ transitions. In an optical experiment\cite{Bertoni2015}, a MC is observed using the probe light at $760$ nm ($1.63$ eV), close to the energy of \textbf{MC2}. Other optical experiments\cite{Hauser1986,Hauser1991,Romstedt1998} demonstrate that there is an obvious MC absorption band ranging from $11000$ \ce{cm^{-1}} ($1.3$ eV) to $14000$ \ce{cm^{-1}} ($1.7$ eV). These means our calculation based on PBE0/6-311G(d,p) is reliable to study the MC of the considered molecule as well.

To well describe the MC in the experiment, the effect of couplings with the STM substrate and that of bias voltages should be similarly taken into account. It has been proven when the high-spin molecule is absorbed on the STM substrate [CuN/Cu(001)], its electronic structure is quite close to that in the free (gas) phase\cite{Gueddida2013}. Hence, the free molecule is still a good model for simulating excitations of the high-spin molecule in the STM experiment. Moreover, the BISST from HS to LS is observed at positive bias in the experiment, and thus we apply the same $\vec{E}^{\texttt{+z}}_{\texttt{UF}}$ to the molecule for simulating the effect of the positive bias (\emph{vide supra}). Fig. \ref{fig4} (a) displays the energy of \textbf{MC1} is $1.4$ eV with $\vec{E}^{\texttt{+z}}_{\texttt{UF}}$ applied, close to that ($1.3$ eV) without $\vec{E}^{\texttt{+z}}_{\texttt{UF}}$. The little influenced energy of \textbf{MC1} is because the $d$-$d$ transition just occurs on the metal ion and the induced transition dipole moment is ignorable. The NTO analysis of \textbf{MC1} in Fig. \ref{fig4} (b) reveals that orbitals of ligands mixed into \textbf{MC1} are greatly reduced in both hole and particle NTOs, and the $d$-$d$ transition is purified at positive bias. The above result means the voltage threshold for activating a MC is about $1.4$ V, consistent with the experiment that a positive bias voltage of $1.5$ V induces the BISST from HS to LS. In the experiment, it is also found the positive bias voltage inducing the BISST from HS to LS can decrease to $1.2$ V, less than our calculated voltage threshold. To understand this bias voltage, we note an experiment in which hot electrons injected from the STM tip are laterally transported via surface resonances at the metallic surface and the dissociation of the molecule is observed\cite{Maksymovych2007}. This dissociation effect is highly possible in the BISST experiment\cite{Miyamachi2012}, since the electron tunneling is similarly from the STM tip to the substrate at positive bias [Fig. \ref{fig2} (a)]. Due to the dissociation effect, the energy gap is decreased between \ce{t_{2g}} and \ce{e_g} orbitals, and the $d$-$d$ transition energy is reduced from $1.4$ V to a lower value. In short, the voltage threshold for activating a MC agrees well with bias voltages used to observe the BISST from HS to LS in the experiment.

For effective access to the MC in the experiment, the applied voltage should be equal to or larger than a related voltage threshold, which is just one hand. The other hand is associated with the bias polarity. When $\vec{E}^{\texttt{-z}}_{\texttt{UF}}$ (simulating $V^{\texttt{-z}}_{\texttt{SD}}$) is applied to the considered molecule, the energy of \textbf{MC1} is $1.5$ eV and the lowest MLCT of the high-spin molecule (\textbf{MLCT4}) is reduced from $1.5$ eV to $1.1$ eV [see Fig. \ref{fig3} (a) and Fig. \ref{fig4} (a)]. The MLCT feature of \textbf{MLCT4} is confirmed by analyzing NTOs in Fig. \ref{fig4}(c), and the MC feature of \textbf{MC1} at negative bias is similar to that at positive bias [Fig. \ref{fig4} (b)], thus it is not shown here. It indicates not only \textbf{MC1} but also \textbf{MLCT4} are accessible at negative bias ($1.5$ V) in the experiment. Nonetheless, the oscillator strength of \textbf{MLCT4} ($f=0.02$) is greatly higher than that of \textbf{MC1} ($f=0.002$), which is also confirmed by experiments\cite{Hauser}. Therefore, \textbf{MLCT4} is predominantly observed at negative bias although both MC and MLCT are accessible. When $\vec{E}^{\texttt{+z}}_{\texttt{UF}}$ (simulating $V^{\texttt{+z}}_{\texttt{SD}}$) is applied to the molecule, the energy of \textbf{MC1} can be reduced to $1.2$ eV in the experiment, but the energy of \textbf{MLCT4} is as high as $1.7$ eV. Only \textbf{MC1} can be accessed by the positive bias ($1.5$ V) in the experiment. The above results illustrate the effective access to the MC is dependent on the bias polarity, which can explain the bias-polarity dependence of the BISST from HS to LS in the experiment.

Finally, based on our present study one could expect that the SST is achieved fast by applying bias voltages in molecular junctions. The fast BISST is due to the SST caused by MLCT and MC, similar to the LIESST. It has been revealed that the SST is observed less than one picosecond following these excitations\cite{Bertoni2015}. Additionally, one could anticipate that the bias-polarity dependence of the BISST (LS$\rightleftharpoons$HS) is easily observed using spin crossover molecules with two different kinds of ligands surrounding the metal ion. Using \ce{Fe(bpz)2(phen)} [bpz=dihydrobis(pyrazoly)borate], \citet{Gopakumar2012} have demonstrated the bias-polarity dependence of the BISST from LS to HS as well.

\section{Conclusions}

In this paper, we propose that the BISST from LS to HS (or from HS to LS) can be accomplished by MLCT (or MC). This proposal is based on the two facts well established in experiments: i) the SST from LS to HS (from HS to LS) is observed following the light-induced MLCT (MC); ii) electronic excitations including the MLCT and MC can be activated by inelastic electron cotunneling in molecular junctions with bias voltages applied. To show the rationality of our proposal, TD-DFT calculations are performed. The calculated voltage threshold for stimulating the MLCT (MC) is consistent with bias voltages used to observe the BISST from LS to HS (from HS to LS) in the experiment reported by \citet{Miyamachi2012}; the realization of MLCT (MC) is determined by the bias polarity, based on which one can well comprehend the bias-polarity dependence of the BISST between LS and HS in the experiment. Our present work helps to design of molcular devices with fast BISST and the special bias-polarity dependence.

\section*{Acknowledgement}

This work was supported by the National Science Foundation of China under Grant Nos. 11374301, 11774349, 11574318, 11534012, and Director Grants of CASHIPS. Calculations were performed partly at the Center for Computational Science of CASHIPS and partly at the Center for Research Computing at the University of Pittsburgh and the Extreme Science and Engineering Discovery Environment (XSEDE).


%
%

%


\bibliography{paper}

\end{document}